\documentclass[prl,twocolumn,amsfonts,showpacs]{revtex4}

\usepackage{amsmath}
\usepackage{graphicx}

\def\be{\begin{equation}}
\def\ee{\end{equation}}
\def \bea#1\eea {\begin{eqnarray}#1\end{eqnarray}}

\def\br{{\bf r}}
\def\bq{{\bf q}}
\def\eps{\varepsilon}
\newcommand{\corr}[1]{\langle #1\rangle}

\def\Gi{{\rm Gi}}
\def\gcf{g_{cF}}

\begin{document}

\title{Superconductivity in disordered thin films: giant mesoscopic fluctuations}

\author{M. A. Skvortsov}
\email{skvor@itp.ac.ru}
\affiliation{L. D. Landau Institute for Theoretical Physics, Moscow 119334, Russia}
\author{M. V. Feigel'man}
\affiliation{L. D. Landau Institute for Theoretical Physics, Moscow 119334, Russia}

\date{March 31, 2005}

\begin{abstract}
We discuss intrinsic inhomogeneities of superconductive properties
of uniformly disordered thin films with large dimensionless conductance $g$.
It is shown that mesoscopic fluctuations, which usually contain a small
factor $1/g$, are crucially enhanced near the critical conductance $\gcf\gg1$
where superconductivity is destroyed at $T=0$ due to Coulomb suppression
of the Cooper attraction.
This leads to strong spatial fluctuations of the local transition temperature
and thus to percolative nature of the thermal superconductive transition.
\end{abstract}


\maketitle

Since the very early stage of superconductivity
theory it is known~\cite{AbrikosovGorkov,Anderson59} that
the superconducting transition temperature, $T_c$,
is insensitive to the rate $\tau^{-1}$ of
elastic impurity scattering,
i.e.\ it does not depend on the parameter $\tau T_c/\hbar$.
This statement known as the ``Anderson theorem''
is valid provided that both (i) Coulomb interaction
effects and (ii) mesoscopic fluctuations are negligible.
However, in sufficiently disordered metals,
close to the Anderson localization transition,
these effects become important and the Anderson theorem is violated.

In disordered samples, Coulomb repulsion
enhanced due to diffusive character of electron motion \cite{AAL80}
leads to the suppression of $T_c$ with the increase of disorder
(cf.\ Ref.~\onlinecite{Fin94} for a review).
For two-dimensional (2D) thin films with the dimensionless conductance
$g=2\pi\hbar/e^2R_{\Box}\gg1$, the first-order
perturbative correction had been calculated in
Refs.~\onlinecite{Japaneese82}, and the general expression
for $T_c(g)$ was obtained by Finkelstein \cite{Finkelstein87}.
In the leading order over $g^{-1/2}$ his result reads:
\be
\label{Tc-Fin}
   \frac{T_c\tau_*}{\hbar}
   =
   \left[
     \frac{\sqrt{2\pi g}-\ln(\hbar/T_{c0}\tau_*)}
          {\sqrt{2\pi g}+\ln(\hbar/T_{c0}\tau_*)}
   \right]^{\sqrt{\pi g/2}} ,
\ee
where $T_{c0}$ is the transition temperature in the clean ($g\to\infty$)
system, and $\tau_* = \max\{\tau, \tau (d/l)^2\}$,
with $d$ being the film thickness and $l=v_F\tau$ being the mean free path.
According to Eq.~(\ref{Tc-Fin}), $T_c$ vanishes at the critical
conductance $\gcf=\ln^2(\hbar/T_{c0}\tau_*)/(2\pi)$
(which is supposed to be large enough for the theory to be self-consistent).

The Finkelstein's theory is an extended version of the
mean-field theory of superconductivity
which takes into account that the effective
Cooper attraction acquires (due to Coulomb interaction and slow diffusion) a
negative energy-dependent contribution.  Within this (``fermionic") mechanism,
vanishing of $T_c$ is accompanied by vanishing of the {\it amplitude}\/ of the
superconductive order parameter $\Delta$.
Another (``bosonic") mechanism of $T_c$ suppression~\cite{Larkin99}
is due to {\it phase fluctuations}\/ of the order parameter.
This mechanism  seems to be adequate mainly for structurally inhomogeneous
superconductors (granular films of artificial arrays)
with well-defined superconductive grains interconnected by weak links.
Below we will see, however, that phase fluctuations inevitably
become relevant for homogeneously disordered films with
the conductance $g$ close to its critical value $\gcf$.

Another  phenomenon important for 2D  conductors is known as
{\it mesoscopic fluctuations}~\cite{UCF}
and is due to nonlocal interference
of electron waves scattering on impurities.
It was recognized by Spivak and Zhou~\cite{SpivakZhou95}
that similar fluctuations
are pertinent also for the Cooper paring susceptibility,
$K(\br,\br') = \corr{K(\br-\br')}+\delta K(\br,\br')$,
which enters the BCS self-consistent equation
\be
  \Delta(\br)
   = \frac{\lambda}{\nu} \int K(\br,\br') \, \Delta(\br') \, d\br' ,
   \label{lin-gap-eq}
\ee
where $\lambda$ is the dimensionless Cooper coupling constant
and $\nu$ is the single-particle density of states per spin.
Equation~(\ref{lin-gap-eq}) with the {\em exact}\/
disorder-dependent kernel $K(\br,\br')$ possesses
localized solutions for $\Delta(\br)$ {\em above}\/
the mean-field transition line. They describe
droplets of superconducting phase which nucleate
prior to the transition of the whole system.
Since the relative magnitude of mesoscopic fluctuations
of $\delta K(\br,\br')$ is of the order of $1/g$ and
is small for a good metal, the effect of localized
droplets on the {\em zero-field}\/ superconductive transition
is negligible and the transition width is determined by
thermal fluctuations.
Contrary, at low temperatures near the upper critical field $H_{c2}(0)$,
thermal degrees of freedom are frozen out and
mesoscopic fluctuations are fully responsible
for the width of the {\em field-driven}\/ superconductor--normal-metal (SN)
transition~\cite{SpivakZhou95,GalitskiiLarkin01}.
Still, the relative magnitude of $H_{c2}(0)$ shift and
of the transition width due to mesoscopic fluctuations
are of the order of $1/g \ll 1$ as long as Coulomb effects are neglected.

The goal of the present Letter is to develop
a combined theory of the superconductive
transition in 2D disordered films,
which takes into account both Coulomb effects and mesoscopic fluctuations.
Our main result is the expression for the relative
smearing $\delta_d=\delta T_c/T_c$ of the
zero-field transition due to formation of localized islands:
\be
   \delta_d = \frac{a_d}{g(g-\gcf)} ,
\label{result}
\ee
where $a_d\approx0.4$.
In the vicinity of the Finkelstein's critical point,
at $g-\gcf\lesssim1$,
formation of islands dominates over thermal fluctuations
characterized by the Ginzburg number
${\rm Gi}=(\pi/8g)$ \cite{AslamazovLarkin68}.
Moreover, in the very close vicinity of the quantum critical point,
at $g-\gcf \lesssim 1/\gcf$, fluctuations of the
``local transition temperature" become large on the absolute scale,
$\delta T_c \sim T_c$, and the superconductive state becomes
strongly inhomogeneous {\it in the absence of any pre-determined structural
granularity}~\cite{KoZvi}.

We emphasize that mesoscopic fluctuations are minimal fluctuations
which are inevitably present in any disordered system.
In real samples, their effect may be enhanced by various
types of structural inhomogeneities.

\begin{figure}
\includegraphics{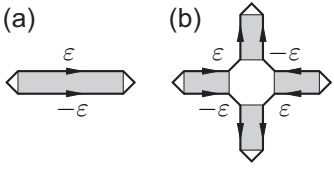}
\caption{\label{F:GLdiagrams} Diagrams for the GL free energy:
(a) $|\Delta|^2$ term;
(b) $|\Delta|^4$ term; its central part is the Hikami box.}
\end{figure}

{\em Ginzburg-Landau expansion}.---%
We begin with deriving the Ginzburg-Landau (GL) expansion
in the vicinity of the Finkelstein transition temperature (\ref{Tc-Fin}).
The GL free energy for the static order parameter
has the form:
\be
   {\cal F}[\Delta]
   =
   \int \left(
   \alpha |\Delta|^2
   + \frac\beta2 |\Delta|^4
   + \gamma |\nabla\Delta|^2
   \right) d\br
   + \tilde{\cal F}[\Delta] ,
\label{GL}
\ee
where the first term is the disorder-averaged contribution,
and the last term accounts for mesoscopic fluctuations;
its form will be found later.
The diagrams for the disorder-averaged free energy~\cite{Gorkov59}
are shown in Fig.~\ref{F:GLdiagrams}. Apart from the standard
impurity averaging, cooperons should be averaged
over fluctuations of the electric field
(this is shown by the gray rectangle).
The order parameter always enters in the combination
$\corr{K_{\eps}(\br-\br')}\Delta(\br')$, where we introduced
the reduced Cooper kernel:
\be
    \corr{K_{\eps_k}(\br-\br')}
    =
    T \sum_{m}
    \corr{G_{\eps_k,\eps_m}(\br,\br') G_{-\eps_k,-\eps_m}(\br,\br')} ,
\label{K-def}
\ee
with $G_{\eps_k,\eps_m}(\br,\br')$ being the Matsubara exact Green function
which, in the presence of the fluctuating electric field, depends on two
energy arguments.
After summation over energy Eq.~(\ref{K-def})
gives the (averaged) pairing susceptibility:
$\corr{K(\br-\br')}=T\sum_{\eps}\corr{K_{\eps}(\br-\br')}$.

\begin{figure}
\includegraphics{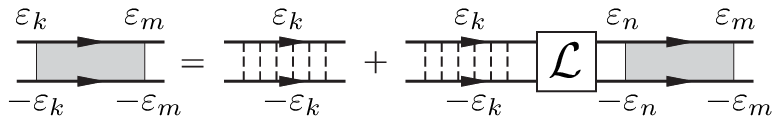}
\caption{\label{F:screening}
Equation for the cooperon in the fluctuating electric field.
The Coulomb vertex ${\cal L}$ is given by Eq.~(\ref{L-vertex}).
The kernel $\corr{K_{\eps_k}(\br-\br')}$
is obtained after summation over $\eps_m$.}
\end{figure}

The kernel $\corr{K_{\eps}(\br-\br')}$
obeys the linear equation shown
diagrammatically in Fig.~\ref{F:screening}.
To write it in a compact form we define
the {\em cooperon screening factor}\/
$w_{\bq}(\eps)$:
\be
   \corr{K_{\eps}(\bq)}
   =
   \frac{2\pi\nu}{Dq^2+2|\eps|} \, w_{\bq}(\eps) ,
\ee
which shows how the free metallic cooperon
gets modified by the fluctuating electric field
($D$ is the diffusion coefficient).
The screening factor satisfies the equation
\be
    w_{\bq}(\eps_k)
    =
    1
    -
    T \sum_{n}
    \frac{{\cal L}_{\eps_k\eps_n}w_{\bq}(\eps_n)}{Dq^2+2|\eps_n|} ,
\label{w-eq}
\ee
where (for thick films with $d>l$ replace $\tau$ by $\tau_*$)
\be
   {\cal L}_{\eps_k\eps_n}
   = \frac{2}{g} \, \theta(\eps_k\eps_n) \, \ln\frac{1}{|\eps_k+\eps_n|\tau}
\label{L-vertex}
\ee
is the disorder-enhanced Coulomb interaction vertex~\cite{OregFin99}
proportional to the return probability,
with $\theta(x)$ being a unit step function.

Equation~(\ref{w-eq}) looks pretty similar to the equation for the
energy-dependent Cooper vertex $\Gamma_{\eps,\eps'}$
considered in Ref.~\onlinecite{OregFin99}.
It can be easily solved with logarithmic
accuracy \cite{OregFin99}.
Introducing a new variable $\zeta=\ln(1/\eps\tau)$,
one finds for the zero-momentum limit of the screening
factor \cite{2B}:
\be
    w(\zeta) \equiv w_0(\eps)
    =
    \cosh(\lambda_g\zeta)
    -
    \tanh(\lambda_g\zeta_T)
    \sinh(\lambda_g\zeta) ,
\label{w(zeta)}
\ee
where $\zeta_T\equiv\ln(1/T\tau)$, and $\lambda_g=1/\sqrt{2\pi g}$
is the Finkelstein's fixed point.
The function $w_0(T)$ decreases from 1 at $\eps\sim1/\tau$
down to $w(\zeta_T)=1/\cosh(\lambda_g\zeta_T)$ at $\eps\sim T$.

The coefficients in the GL free energy (\ref{GL}) are given by:%
\begin{subequations}%
\label{GL-coefficients}%
\begin{gather}
   \frac{\alpha}{\nu}
   = \frac{1}{\lambda} - \pi T \sum_\eps \frac{w_0(\eps)}{|\eps|}
   = \frac{1}{\lambda_*} - \int_0^{\zeta_T} d\zeta \, w(\zeta) ,
\label{F2}
\\
   \gamma
   = \frac{\pi\nu D T}{2} \sum_\eps \frac{w_0^2(\eps)}{\eps^2}
   = \gamma_0 w^2(\zeta_T) ,
\label{F2grad}
\\
   \beta
   = \frac{\pi\nu}{2} \sum_{\eps} \frac{w_0^4(\eps)}{|\eps|^3}
   = \beta_0 w^4(\zeta_T) ,
\label{F4}
\end{gather}
\end{subequations}
where $\beta_0 = 7\zeta(3)\nu/(8\pi^2T_c^2)$
and $\gamma_0 = \pi\nu D/(8 T_c)$ are the standard
coefficients for dirty superconductors \cite{Gorkov59},
and $\lambda_*$ is the running Cooper coupling constant
at the energy scale $\tau^{-1}$.
The Matsubara sums in Eqs.~(\ref{F2grad}) and (\ref{F4})
converge at the thermal scale.
Therefore the coefficients $\beta$ and $\gamma$ contain
the screening factors evaluated at $\zeta_T$.
Contrary, the coefficient $\alpha$ is determined
by all energies $\eps<1/\tau$.
The integral in Eq.~(\ref{F2}) is given by
$\lambda_g^{-1} \tanh (\lambda_g\zeta_T)$,
and solving $\alpha(T_c)=0$ one immediately recovers
the Finkel'stein expression (\ref{Tc-Fin}) for $T_c$.
Taking the derivative near $T_c$
we find $\alpha=\nu(T/T_c-1) w^2(\zeta_T)$.

Thus, disregarding mesoscopic fluctuations, we see that
the $\Delta$ field always enters the GL expansion
in the combination with the screening factor
$w(\zeta_T)=1/\cosh(\lambda_g\zeta_T)$.
If we define $\tilde\Delta=\Delta w(\zeta_{T_c})$
then the GL expansion for $\tilde\Delta$ will acquire
the standard form with the coefficients
$\alpha_0=\nu(T/T_c-1)$, $\beta_0$ and $\gamma_0$.
As a corollary, the Ginzburg number appears to be unaffected
by the Coulomb repulsion: $\Gi=\pi/(8g)$.

{\em Quasiparticle spectrum}.---%
Incorporating the Coulomb screening factors $w_\bq(\eps)$
into the standard Green function formalism, one finds
that a quasiparticle propagating with the energy $\eps$
feels an effective pairing potential
$\Delta_{\text{eff}}(\eps)= w(\zeta)\Delta$.
The gap in the spectrum, $E_{\text{gap}}$,
should be found as a solution
to $E_{\text{gap}}=w[\ln(1/E_{\text{gap}}\tau)]\Delta$.
Deep in the superconducting phase, at $T_c-T\sim T_c$,
it coincides with $\tilde\Delta$.
On the other hand, excited particles with $\eps>E_{\text{gap}}$
see a larger value of $\Delta_{\text{eff}}(\eps)$,
and high-energy particles with $\eps\gtrsim1/\tau$
feel the bare value of $\Delta$.

In the regime of strong Coulomb suppression of superconductivity,
when $T_c\ll T_{c0}$ and $w(\zeta_{T_c})\ll1$,
the bare $\Delta$ significantly exceeds the screened $\tilde\Delta$.
This enhancement of $\Delta$ was irrelevant for the GL expansion
since the sums over Matsubara energies which determined
the GL coefficients converged at the thermal scale.
We will see below that this is not the case for mesoscopic
fluctuations where the bare value of $\Delta$ comes into play.

\begin{figure}
\includegraphics{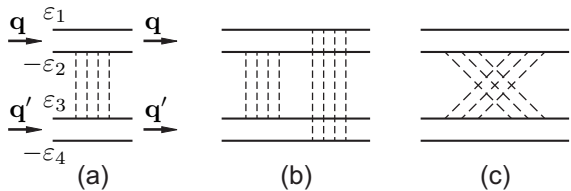}
\caption{\label{F:mesofluct} One-loop diagrams for
the 4-cooperon/diffuson collision vertex $M_{\eps_i}(\bq,\bq')$.
The diagrams (a) and (c) have their symmetric counterparts.}
\end{figure}

{\em Mesoscopic fluctuations of the pairing susceptibility}.---%
In order to calculate the correlation function
of the pairing susceptibility, one has to draw
two diagrams for $K(\br,\br')$ (see Fig.~\ref{F:GLdiagrams}a)
and connect their diffusive modes by impurity lines.
In general,
the variance $\corr{{\delta K(\br_1,\br_2)}{\delta K(\br_3,\br_4)}}$
is a complicated function of $\br_i-\br_j$ decaying at the scale
of the thermal length $L_T=\sqrt{D/(2\pi T)}$. On the other hand,
close to $T_c$ the order parameter varies at the scale
of the coherence length $\xi(T)= L_T\sqrt{T_c/(T-T_c)}\gg L_T$.
Therefore, in the vicinity of the superconducting transition
the fluctuations of $K(\br,\br')$ are effectively {\em short-ranged}\/
and characterized by the single number
\be
   C
   = \int \corr{{\delta K(\br_1,\br_2)}{\delta K(\br_3,\br_4)}}
     \, d\br_2 d\br_3 d\br_4 .
\label{C-def}
\ee
Taking into account all possible correlations between
diffusive modes in $K(\br,\br')$ we find
\begin{align}
   C
   =
   T^4
   \sum_{\eps_i>0}
   \left(
     \prod_{i=1}^4 \frac{w_0(\eps_i)}{\eps_i}
   \right)
   \hat{\cal R}^{12}_{\bq} \,
   \hat{\cal R}^{34}_{\bq'} \,
   M_{\eps_i}(\bq,\bq') ,
\label{C-long}
\end{align}
where $\hat{\cal R}^{ij}_{\bq}$ is an operator acting
on an arbitrary function $X(\bq)$ as
\be
   \hat{\cal R}^{ij}_{\bq} \, X(\bq)
   =
   \frac{\delta_{\eps_i\eps_j}}{2T} X(0)
     + \frac{1}{\nu}
       \int \frac{d\bq/(2\pi)^2} {[D\bq^2+\eps_{ij}]^2} X(\bq) ,
\label{R-oper}
\ee
$\eps_{ij}\equiv\eps_i+\eps_j$,
and $M_{\eps_i}(\bq,\bq')$ is the 4-cooperon/diffuson
collision vertex shown in Fig.~\ref{F:mesofluct}
(with the proper construction of internal Hikami boxes by drawing
additional impurity lines being implied).
The internal diffusive modes can be diffusons or cooperons.
In the vicinity of $H_{c2}(0)$,
the diagram (b) was considered in Refs.~\onlinecite{SpivakZhou95},
and the diagram (c) was analyzed in Ref.~\onlinecite{Lamacraft04}.
At zero magnetic field the vertex is calculated elsewhere~\cite{2B}:
\be
   M_{\eps_i}(\bq,\bq')
   = \frac{[D\bq^2+\eps_{12}][D\bq'^2+\eps_{34}]}{2\pi D}
     \frac{\eps_{14}+\eps_{23}}{\eps_{14}\eps_{23}} .
\label{M}
\ee

The first term in Eq.~(\ref{R-oper})
refers to the cooperons in the ladder (\ref{w-eq}) shown
in Fig.~\ref{F:screening}, while its second term
refers to the cooperons/diffusons
which are responsible for the return probability
in the vertex $\cal L$ given by Eq.~(\ref{L-vertex}).
Some summations over energies in Eq.~(\ref{C-long}) saturate
at the thermal scale, whereas summations associated with return
probability are logarithmic extending up to the high-energy
cutoff $\tau^{-1}$:
\begin{align}
   C
   = &
   \frac{w_0^2(T)}{\pi D}
   \left(
     T^2 \sum_{\eps,\eps'>0} \frac{1}{\eps\eps'(\eps+\eps')}
   \right)
\nonumber
\\
   & {}\times
   \left[
     w_0(T)
     +
     \frac1g
       T \sum_{\eps>0} \frac{w_0(\eps)}{\eps} \ln\frac{1}{\eps\tau}
   \right]^2 .
\label{C2}
\end{align}
The second term in the square brackets in Eq.~(\ref{C2})
is due to mesoscopic fluctuations of return probability.
Calculating the sum as an integral over $\zeta$ with the help
of Eq.~(\ref{w(zeta)}) we find that this term is equal to $1-w_0(T)$,
so that the total expression in the square brackets in Eq.~(\ref{C2})
is equal to 1.
Thus, in the regime of strong Coulomb suppression of superconductivity,
the pairing susceptibility fluctuates mainly due to mesoscopic fluctuations
of return probability in the vertex (\ref{L-vertex}).
Writing $C=C_0w_0^4(T)$, we get
\be
   C_0 = \frac{7\zeta(3)}{8\pi^4DT} \cosh^2(\lambda_g\zeta_T) .
\label{<<KK>>}
\ee

{\em Superconductor with fluctuating $T_c$}.---%
Short-range mesoscopic fluctuations of $\delta K(\br_1,\br_2)$
are equivalent to local fluctuations
of the transition temperature $\delta T_c(\br)$
which can be described by the following term
in the GL free energy:
\be
   \tilde{\cal F}[\tilde{\Delta}] =
     \int \delta\tilde\alpha(\br)
      \, |\tilde\Delta({\bf r})|^2\,
d{\bf r} ,
\ee
where $\corr{\delta\tilde\alpha(\br)\delta\tilde\alpha(\br')}=
C_0\delta(\br-\br')$.
Superconductors with local fluctuations of $T_c$ were considered previously
within the phenomenological approach by Ioffe and Larkin~\cite{IoffeLarkin81},
where the three-dimensional case was mainly discussed.
Generalizing their results to the 2D case,
we find that the relative smearing
of the superconductive transition due to frozen-in mesoscopic fluctuations
is given by $\delta_d = C_0/[12 T_c \gamma_0 (d\alpha_0/dT)]$
(note that in the 2D case the numerical coefficient in the exponent
of Eq.~(29) in Ref.~\onlinecite{IoffeLarkin81} is equal to 11.8,
cf.~\cite{LifshitsGredeskulPastur}).
Taking $C_0$ at $T=T_c$ and using Eq.~(\ref{Tc-Fin}) we obtain
a surprisingly simple expression~(\ref{result}) with $a_d=28\zeta(3)/3\pi^3$.

The increase of $\delta_d$ near the critical conductance $g_{cF}$
can be understood in terms of renormalization of
the Cooper attraction constant $\lambda$.
At low energies, the latter
acquires a negative Coulomb contribution proportional
to the return probablity $\sim g^{-1}\ln(1/\epsilon\tau)$.
Mesoscopic fluctuations of $g$ lead then
to fluctuations of $\lambda$, whose relative effect grows with
decreasing $\lambda$:
$\delta_d \sim \delta T_c/T_c = \delta\lambda/\lambda^2$.

Equation (\ref{result}) predicts that
for $g-\gcf \lesssim 1$, the disorder-induced
broadening of the transition dominates over the thermal width:
$\delta_d>\Gi$.
In such a situation, the macroscopic superconductive transition occurs
via formation of small superconductive islands
of size $L_T=\sqrt{D/(2\pi T_c)}$
surrounded by the normal metal state. With the temperature decrease,
the density of these islands and the proximity-induced coupling between them
grows until a percolation-type superconductive transition~\cite{IoffeLarkin81}
takes place.  At sufficiently low temperatures, $T\lesssim T_c(1-\delta_d)$,
the superconductive state becomes approximately uniform, with weak
spacial variations in the amplitude of the order parameter $|\Delta|$.

Another situation occurs in the closest vicinity of the critical conductance,
$g - \gcf \lesssim 1/\gcf$: here local fluctuations of $T_c$
are large on the absolute scale, and strong inhomogeneity
of the superconductive order parameter persists down to  $T \ll T_c$.
As a result, both thermal and quantum fluctuations of phases
of superconductive order parameters on different superconductive
islands are strongly increased. In other terms, in the close vicinity
of the critical conductance $\gcf$,
the {\it bosonic}\/ mechanism of superconductivity suppression becomes relevant.

Inhomogeneous distribution of $|\Delta(\br)|$
is known~\cite{LO72,MeyerSimons01} to smear the gap
in the excitation spectrum of a superconductor. We expect this effect
to be very strong for $g \approx \gcf$.

The result (\ref{result}) indicates that strong enhancement
(in comparison with the results
of Refs.~\onlinecite{SpivakZhou95,GalitskiiLarkin01})
of mesoscopic fluctuation effects in $H_{c2}$ behavior at low temperature
should be expected at $g \approx \gcf$. This problem needs  special
treatment since correlation length of mesoscopic fluctuations diverges at
$T \to 0$, and short-range approximation employed for the
determination of $C$ in Eq.~(\ref{C-def}) becomes
inappropriate~\cite{Lamacraft04}.
It is quite clear however, that
long-range features of mesoscopic disorder at $ T\to 0$ can only increase
the effective width of the field-driven $T=0$
SN transition in comparison with
the width $\delta_d$ of the zero-field transition driven by temperature.
An extension of the present
approach to magnetic-field-induced transition near $H_{c2}(0)$
will be considered separately.

To conclude, we demonstrated that strong inhomogeneities of superconductive
state can be induced by  relatively weak $(\sim g^{-1})$ mesoscopic
fluctuations, which lead to spatial fluctuations of the effective Cooper
attraction constant.
As a result, a nominally uniformly disordered film may appear as a granular
one in terms of its superconductive properties.

We thank B. L. Altshuler and T. I. Baturina for stimulating discussions.
This research was supported by the Program ``Quantum Macrophysics" of
the Russian Academy of Sciences and  RFBR under
grants No.\ 04-02-16348, 04-02-08159 and 04-02-16998.
M.~A.~S. acknowledges support from the Dynasty foundation
and the ICFPM.

\end{document}